\documentclass{article}

\usepackage[english]{babel}

\usepackage[letterpaper,top=2cm,bottom=2cm,left=3cm,right=3cm,marginparwidth=1.75cm]{geometry}

\usepackage{amsmath}
\usepackage{graphicx}
\usepackage[colorlinks=true, allcolors=blue]{hyperref}
\usepackage[authoryear]
{natbib} 
\usepackage{filecontents}
\usepackage{amssymb}
\usepackage{hyperref}
\usepackage{listings}
\usepackage{mathtools}
\usepackage{maple}
\usepackage[utf8]{inputenc}
\usepackage[svgnames]{xcolor}

\usepackage{breqn}
\usepackage{textcomp}

\DefineParaStyle{Maple Bullet Item}
\DefineParaStyle{Maple Heading 1}
\DefineParaStyle{Maple Warning}
\DefineParaStyle{Maple Heading 4}
\DefineParaStyle{Maple Heading 2}
\DefineParaStyle{Maple Heading 3}
\DefineParaStyle{Maple Dash Item}
\DefineParaStyle{Maple Error}
\DefineParaStyle{Maple Title}
\DefineParaStyle{Maple Ordered List 1}
\DefineParaStyle{Maple Text Output}
\DefineParaStyle{Maple Ordered List 2}
\DefineParaStyle{Maple Ordered List 3}
\DefineParaStyle{Maple Normal}
\DefineParaStyle{Maple Ordered List 4}
\DefineParaStyle{Maple Ordered List 5}
\DefineCharStyle{Maple 2D Output}
\DefineCharStyle{Maple 2D Input}
\DefineCharStyle{Maple Maple Input}
\DefineCharStyle{Maple 2D Math}
\DefineCharStyle{Maple Hyperlink}

\def\H{\mbox{I\hspace{-.15em}H\hspace{-.15em}I}}
\def\1{\mbox{I\hspace{-.15em}1}}

\def\b{\begin{equation}}
\def\e{\end{equation}}
\def\bee{\begin{enumerate}}
\def\eee{\end{enumerate}}
\title{The measurement problem in liquid NMR quantum computers}
\author{O. Jalili$^1$\thanks{E-Mail:~omid\_jalili@yahoo.com \& om.jalili@iau.ac.ir}, M. Heidari Adl$^{1}$\thanks{E-Mail:~mn.heidariadl@gmail.com} }
\date{\today}

\begin{document}
  \maketitle {\it \centerline{\it $^1$ Department of Physics, Nour branch
,Islamic Azad University, Nour, Iran }\centerline{\it P.O.BOX
46415/444, Nour, Mazandaran, IRAN} }
\begin{abstract}

With the quantum state diffusion measurement theory(QSD), the measurement problem in liquid nuclear magnetic resonance(NMR) quantum computers was addressed and then it was shown that due to statistical fluctuations, the measured magnetic moment value is comparable to noise for long times. Therefore, we suggested that the measurement time should be short to distinguish signal from noise.
\end{abstract}

%\keywords{NMR quantum computer, Measurement problem, Quantum state diffusion, Continuous measurement}

\vspace{0.5cm} {  \textbf{Keywords}}:\text{~NMR quantum computer, Measurement problem, Quantum state diffusion}  \vspace{0.2cm}    

{\textbf{ Proposed PACS numbers}}: 03:65.-w, 03.65.Ta, 03.65.Yz, 03.67.Lx.      \vspace{0.5cm}

\section{Introduction}
\label{sec:intro}
\begin{figure}[t]
\centering
\includegraphics[width=0.3\textwidth]{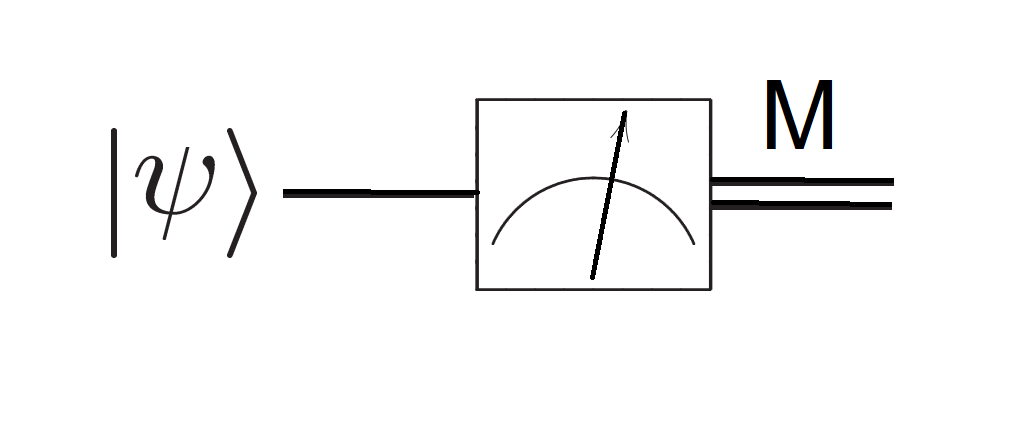}
\caption{\label{fig:meter}Quantum circuit symbol for measurement. The act of measuring is represented by a meter. For distinction, we represent classical beats with two lines.}
\end{figure}

 In every quantum computer, there is a part or parts called measurement, whose circuit symbol is shown in Figure \ref{fig:meter}. In the measurement operation, the qubit is converted into a classical bit. In the measurement process, the coherence of the device is destroyed and the device jumps to one of the eigenstates of the measurement operator, a phenomenon called a quantum jump. Jumps do not only occur in the measurement operation. From the history of quantum mechanics, we know that Einstein assumed a spontaneous transition in the theory of the coefficients A and B, which is a kind of quantum jump. In fact, in the world of quantum mechanics, we have two types of time evolution (dynamics), the "dynamics" that are controlled by the Schrödinger equation and the "dynamics" that result from jumps (which are usually a measurements). Although we know the "dynamics" of the Schrödinger equation well, the "dynamics" of measurement has been a mysterious process: the exact time of the transition is unknown and the structure of the transition is not clear. We only know what the probability of possible events is! Worse still, in the Copenhagen interpretation, an intelligent observer is required for such transitions to occur \cite{RN5,RN3}! That is, in this interpretation, the transitions only occur when an intelligent observer measures them. In this way, the results of the measuring devices remain uncertain until an intelligent observer observes them!

It is clear that such an interpretation is not very satisfactory. In the 1980s and 1990s, attempts were made to explain the jumps caused by measurements solely by Schrödinger dynamics. This interpretation is called quantum state diffusion(QSD) that is a kind of decoherence \cite{RN11,RN12,RN17,RN14,RN13,RN8}. In this theory, the actual measurement is time-consuming, albeit short. During the measurement process, the measuring device interacts with the device being measured. Since measuring devices are generally macroscopic, they have a high degree of freedom. Therefore, the time evolution of the device being measured can be considered as a statistical Schrödinger equation, that is, a Schrödinger equation with statistical terms. In mathematics, these types of equations are called statistical differential equations \cite{RN21,RN23,RN22}. The creators of QSD theory claim that with this formulation, the quantum jumps caused by the measurement process can be explained! The QSD is not the only measurement theory. For example, another theory was developed that recognizes the transitions as a "dynamic" independent of Schrödinger dynamics and assumes that the transitions cannot be extracted from Schrödinger "dynamics". In this theory, at any moment the system can evolve under Schrödinger "dynamics" or based on quantum transitions. Therefore, at the end of a certain period of time, we must average all possible paths to obtain the desired observable value. This method is called the Monte Carlo method or the Monte Carlo wave function method \cite{RN7,RN6,RN8}. There are other interpretations, including the Bohm hidden variable interpretation, parallel universes, etc \cite{RN24}. In the present paper, we have applied the QSD method to the measurement stage in NMR quantum computers and in the end we have come to the conclusion that due to the fluctuations caused by the measurement operation, classical averaging should be performed over a certain time interval. We have also made a recommendation for the averaging period.

The article is as follows: In Section \ref{sec:liquid NMR}, quantum  liquid NMR  computers were introduced and in particular, measurements in this type of computer were introduced. In Section \ref{sec:QSD}, the quantum state diffusion theory was introduced. Quantum state diffusion theory is a theory of quantum measurement that claims that quantum jumps have a kind of stochastic differential equation. Measurement in liquid NMR computers is discussed in Section \ref{sec:Measurements in NMR} in details. In Section \ref{sec:QSD Measurements in NMR}, the problem of measurement in liquid NMR quantum computers is addressed using the QSD theory method. Discussion of the results of this work is postponed to the Conclusion section \ref{sec:Conclu}.

\section{Liquid NMR quantum computers}
\label{sec:liquid NMR}
\begin{figure}[tb]
\centering
\includegraphics[width=0.3\textwidth]{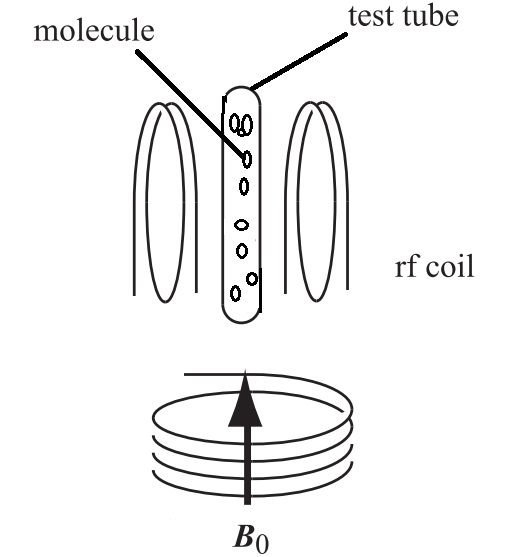}
\caption{\label{fig:NMR}The test tube, which is considered the heart of quantum computers, performs calculations by changing the spin state of the  nuclei of the liquid in the test tube. }
\end{figure}

From Dirac's theory for fermions, we know that the interaction energy of a fermion of charge  $q$ and mass $m$ with a uniform magnetic field $B_{0}$ is equal to
\b H_0=-\frac{q}{mc} S.B_0 \e
where $S$ is the spin. In analogy to the classical interaction of magnetic dipoles with a magnetic field, we rewrite the interaction energy as:
\b H_0=-\mu.B_0. \e
In this case, the magnetic moment is:
\b \mu=\frac{q}{mc} S \e
which can be rewritten as:
\b \mu=\gamma S. \e
$\gamma=\frac{q}{mc}$ is called the gyromagnetic ratio. If we take the z-axis direction to be in the direction of the magnetic field, we have
\b H_0=-\gamma B_0 S_z=\omega_0 S_z \e
where $B_0$ is the magnitude of the magnetic field and $\omega_0=\gamma B_0$ the Larmor frequency. In the case of the nucleus of atoms, which can contain several protons and neutrons, we can still write the relationship $H_0=-\gamma B_0 S_z$, but in this case the following rule should be kept in mind \cite{RN4}
\begin{itemize}
\item If the number of neutrons and the number of protons are both even, the
nucleus has no spin,
\item If the number of neutrons plus the number of protons is odd, then the
nucleus has a half-integer spin (i.e., 1/2, 3/2, 5/2),
\item If the number of neutrons and the number of protons are both odd, then
the nucleus has an integer spin (i.e., 1, 2, 3).
\end{itemize}
In addition, the gyromagnetic ratio also depends on the number of nuclei. We consider the case where the total spin is one-half. In this case, the state space is two-dimensional and the energy of the device is $E_0=-\hbar\omega_0/2$ and $E_1=\hbar\omega_0/2$. The eigenstates of the energy states are:
\b   |1 \rangle=\left( \begin{array}{clcr} 0  \\ 1 \\    \end{array} \right)
  ,|0 \rangle=\left( \begin{array}{clcr} 1 \\ 0 \\  \end{array} \right).\e
In NMR quantum computers, these same nuclear spin states are used as qubit registers \cite{RN20,RN19}. These nuclei, when exposed to a fixed magnetic field, will have a precessional motion around the magnetic field with an angular frequency of $\omega_0$. To control the spin state of the nucleus, a time-varying magnetic field is used. These variable fields are applied in a direction perpendicular to the direction of the fixed field (for example, the x direction ):
\b \textbf{B}_1 (t)=-B_1 (t)\cos(\omega_{rf} t-\phi) \widehat{x}\e
where $B_1 (t)$ is the envelope of the radio frequency field (in the microwave range) which is very slowly varying compared to the frequency $\omega_{rf}$. In this case, the total Hamiltonian is:
\b H=H_0+H_{rf}=-\omega_0S_z+2\omega_1 \cos(\omega_{rf} t-\phi) S_x\e
where $2\omega_1=\gamma B_1$ . The parameters $B_1, \phi, \omega_{rf}$ and the pulse shape are our control parameters through which we can change the state of the device as desired.
In NMR quantum computers, molecules with spin-1/2 nuclei are used that are liquid at room temperature, such as chloroform, cytosine, trichloroethylene, etc. Depending on the molecule, we have one, two or more nuclei with spin-1/2, which can be the same or different. Even in the case of identical nuclei, due to the location of the nucleus in the molecule, the gyromagnetic ratio is generally different for these nuclei, and hence these nuclei can be addressed. The general scheme of the heart of such a computer is shown in the Figure \ref{fig:NMR}.
 It is clear from the nature of the system that the measurement in NMR quantum computers have statistical nature. Because the Hamiltonian of the system is time-varying, this phenomenon is not in equilibrium and the Maxwell-Boltzmann distribution cannot be used. So to find the density function, the Liouville equation must be solved:
\b i\hbar \frac{\partial\widehat{ \rho}}{\partial t}  =[\widehat{H},\widehat{\rho}]\e
To solve this equation,  instead of the laboratory device,  it is better to go to a rotating frame that rotates around the magnetic field with an angular velocity equal to the Larmor frequency. In this case, by choosing $\omega_{rf}$ equal to the Larmor frequency $\omega_0$, the Hamiltonian in the rotating device will be as follows:
\b \widetilde{H} \cong \omega_1(\cos\phi S_x+\sin\phi S_y) \e
This Hamiltonian is traceless, so it is only the generator of SU(2) elements. By choosing the control parameter $\omega_{rf}$ equal to the Larmor frequency, of the four control parameters, three of them now remain: $\omega_1, \phi$, and the pulse shape. The time evolution operator in the rotating system now is:
\b U(t)=e^{-i\int_{0}^{\tau}\widetilde{H}dt}=e^{-i\omega_1 \tau(\cos\phi S_x+\sin\phi S_y) }\e
It can be shown that by appropriately choosing the remaining control parameters, we can easily construct the single gates $X, Y$ and $Z$ \cite{RN19}:
\b X=e^{-i(\pi/2\hbar)S_x}=\frac{1}{\sqrt{2}}\left( \begin{array}{clcr} 1 & -i \\ -i & ~~1 \\ \end{array} \right), Y=e^{-i(\pi/2\hbar)S_Y}=\frac{1}{\sqrt{2}}\left( \begin{array}{clcr} 1 & -1 \\ -1 & ~~1 \\ \end{array} \right)$$ $$,Z=e^{-i(\pi/2\hbar)S_Z}=\frac{1}{\sqrt{2}}\left( \begin{array}{clcr} 1-i & 0 \\ 0 & 1+i \\ \end{array} \right).\e
For example, to create $X$, it is sufficient that $\phi=0$ and the envelope shape of the radio frequency field is a rectangular pulse of length $\tau$ and the relation $\omega_1 \tau=\frac{\pi}{2}$ holds.
Having single gates $X, Y$ and $Z$, any desired single gate can be created. Therefore, starting from a given state, then applying an appropriate sequence of single gates $X, Y$ and $Z$, we can create any desired single gate. Finally, having the interaction between spins, we can also create two-qubit gates.

After the desired state is created with a sequence of single gates X, Y and Z, it is time to measure. As mentioned earlier, the measurement in NMR quantum computers is in the form of ensemble average. The measured quantity is the magnetic moment of the material inside the test tube. For example, the magnetic moment in the x direction is:

\b M_x \propto\langle I_x\rangle=\textbf{tr}(I_x\rho)\e

where $I_x=\frac{\sigma_x}{2}$ and $\sigma_x$ is the Pauli matrix. This work involves some points that we will raise in section \ref{sec:Measurements in NMR}. But now we turn to the theory of quantum state diffusion.
\section{The quantum state diffusion(QSD)}
\label{sec:QSD}

In QSD theory, the measurement act is exactly an interaction. Therefore, like any interaction, it assumes an interaction Hamiltonian for the measurement operation. Therefore, the system under study is not isolated. We assume that the total system, that is, the system being measured and the measuring device, are isolated. Therefore, we do not have quantum jumps in the total system. In this case, we can show that for each subsystem, for example, the measured system, The density matrix applies to an equation called the master equation \cite{RN26}. The Master equation is a generalization of the Liouille equation for open systems. In 1976, the Swedish mathematical physicist Lindelblad showed that the Master equation can be written as follows \cite{RN10}(see \cite{RN19} for a nice proof of it).
\b \dot{\rho} =-\frac{i}{\hbar}[H,\rho]+\Sigma_{j}(L_j \rho L_j^{\dag}- \frac{1}{2}L_j^{\dag}L_j\rho- \frac{1}{2}\rho L_j^{\dag}L_j)\e
where the operator $L_j$ may or may not be Hermitian. The operator $L_j$ is called the Lindenblad operator. Lindenblad operators are determined by the nature of the environment.

The master equation is the governing equation for the density, not the governing equation for the wave function itself. But in the QSD theory, instead of the governing equation for the density matrix, an attempt is made to obtain the governing equation for the state function of the system being measured. This was historically due to computational considerations. Because in the master equation, if the dimension of space is $n$, the number of variables of the density matrix is $n^{2}$, while the number of variables of the state function is $n$ \cite{RN9}. In the QSD, we pay attention to the transformation of the wave function itself and we seek to obtain quantum jumps solely by the Schrödinger transformation \cite{RN17}.

For this purpose, in the QSD theory, the time evolution at any short instant $dt$ is due to two factors: first, the "dynamics" of the system itself due to its own Hamiltonian and second, due to the interaction of the particles of the measuring system on the measured system. But the measuring system is much larger than the measured system, that is, the degree of freedom of the measuring system is much greater than the degree of freedom of the measured system. Also, the effect of each of these degrees of freedom on the observed system is very small so that the conditions of the central limit theorem hold. With this description, we can mathematically examine the problem of "dynamics" of open system as follows. Suppose that the one-dimensional dynamic variable $x(t)$ has the following structural equation.
\b \frac{dx}{dt}=a(t,x) \e
where $a(t,x)$ is a function of time and $x(t)$ is a function of time. This system is assumed to be isolated, that is, it is assumed that no statistical phenomena are present. Now we introduce statistical phenomena. If we assume the system to be open and under the influence of a large system so that the conditions of the central limit theorem hold, then we can write the above equation as follows.
\b \frac{d}{dt} X_t=a(t,X_t )+b(t,X_t ) \xi_t \label{eq:SDE0}\e
Where $b(t,X_t )$ is a bivariate function and $\xi_t$ is also a random variable with a Gaussian distribution (we know this from the central limit theorem). Here, we have shown the function under consideration with $X_t$  to distinguish it from the non-statistical situation, where the index is the function argument. The first term of the expression ~(\ref{eq:SDE0}) is called the drift and the second term is called the diffusion, which represents the noise. The $\xi_t$ is a Gaussian-$N(0,1)$ variable that is independent of each other at different times, that is, for $t^{'}\neq t$ we have: $E(\xi_t,\xi_{t^{'}})=0$. The process $\xi_t$ is called white Gaussian noise. It seems that white Gaussian noise can be written as a derivative of the Wiener process. But it turns out that the Wiener process is not differentiable anywhere! Let us temporarily assume that it is differentiable. The Gaussian process $W_t,t>0$ where $E(W_t )=0$ and $E((W_t )^2 )=t$ and the increments are independent, is called Wiener: The Gaussian process $W_t,t>0$ where $E(W_t )=0$ and $E((W_t )^2 )=t$ and the increments are independent is called Wiener:
\b E((W_{t_4 }-W_{t_3 } )(W_{t_2 }-W_{t_1 } ))=0\e
where $0\leq t_1<t_2\leq t_3<t_4$. So we can write the relation ~(\ref{eq:SDE0}) as follows.
\b dX_t=a(t,X_t )dt+b(t,X_t ) dW_t \label{eq:SDE}\e
If this relation is meaningful, its integral is as follows.
\b X_t=X_{t_0 }+\int_{t_0}^{t}a(t,X_t )dt+\int_{t_0}^{t}a(t,X_t )dW_t \e
While the first integral is completely meaningful and is a Riemann sum, the second integral is meaningless, because the expression $dW_t$ is symbolic and we have not yet given a precise definition of it. In the forties, the Japanese mathematician Ito gave a statistical definition of the second integral. Later, in the sixties, the Russian physicist Stratonovich introduced another type of statistical integrals. Each of these definitions has its own characteristics. So now the ~(\ref{eq:SDE}) relation is meaningful. That is, we know this relation from the definition of the statistical integral. In short, the expression $dW_t$ is meaningful and applies like the mathematical differential in Leibniz's rule \cite{RN21,RN23,RN22}.

Now we return to the dynamics of the Schrödinger equation. According to what was said, the time evolution of the state vector of a system that, in addition to the time evolution due to its own Hamiltonian, is affected by environmental noise can be written as follows.
\b |d\psi\rangle=|a\rangle dt+|b\rangle dw \label{eq:SDE1}\e
where the first term is the drift and the second term is the diffusion. If the diffusion term did not exist, $|\psi\rangle$ would hold in the Schrödinger equation. But because of the diffusion term, the system is not isolated and our goal now is to obtain the governing equation for $|\psi\rangle$. For open systems, we know that the governing equation for the density matrix is the master equation. A density operator $\rho$ for an evolving quantum system can be expressed in many ways as a mean $M$ over a distribution of normalized pure state projection operators. We seek differential equations for $|\psi\rangle$ such that the density operator given by the ensemble mean over the projectors \cite{RN17}.
\b \rho=M(|\psi\rangle \langle\psi|\e
 Therefore, using the stochastic growth of ~(\ref{eq:SDE1}), we write the density matrix changes $\rho=M(|\psi\rangle \langle\psi|)$, as follows:
\b d\rho=M(|d\psi\rangle \langle\psi|+|\psi\rangle \langle d\psi|+|d\psi\rangle \langle d\psi|) \e
By substituting the equation ~(\ref{eq:SDE1}) into the above relation and matching it with the master equation, we can obtain the coefficients $|a\rangle$ and $|b\rangle$, in which case the equation ~(\ref{eq:SDE1}) becomes as follows \cite{RN17}:
\b |d\psi\rangle =-\frac{i}{\hbar} H|\psi\rangle dt+\Sigma_j (\langle L_j^{\dag} \rangle L_j-\frac{1}{2} L_j^{\dag} L_j-\frac{1}{2}\langle L_j^{\dag} \rangle \langle L_j \rangle )|\psi\rangle dt+\Sigma_j(L_j-\langle  L_j \rangle)  |\psi\rangle dw \label{eq:QSD0}\e
This equation is the result we wanted. But because the averages are with respect to $|\psi\rangle$, this equation is strongly nonlinear. However, numerically we can obtain the value of the function at any desired time, from which we can obtain the value of the observables at any desired time. Of course, if the system is not pure, then we calculate each of the elements of the density matrix according to the above equation and then we obtain the total density matrix at the desired time. As we said earlier, the Lindenblad operators are obtained from the nature of the environment (the measuring system). Therefore, in the next section we will examine measurements in NMR quantum computers.
\section{Measurements in liquid NMR quantum computers}
\label{sec:Measurements in NMR}
The quantity measured in NMR quantum computers is the magnetic moment. Since the sample is at room temperature and $\hbar\omega_0\ll K_B T$, the initial state of density matrix is:
\b \rho_{th}=\frac{e^{-\frac{H_0}{K_B T}}}{Z(T)}\cong\frac{1}{2} (1+\frac{\hbar\omega_0}{K_B T} I_z ), Z(T)=\textbf{tr} e^{\frac{-H_0}{K_B T}}\e
where $I_z=\frac{\sigma_z}{2}$ and $\sigma_z$ is the Pauli matrix. This is the result in the laboratory frame. In the rotating frame, we have:
\b \widetilde{\rho}_{th}=U\rho_{th} U^{†}=\rho_{th} \e
where the evolution operator is $U=\exp(i\omega_0 I_z t)$ and the last equality can be shown easily. Now we turn to the quantum operations relevant to our calculations. There are two types of quantum gates, single-qubit and double-qubit. Since our purpose is to study measurement theory, we will limit ourselves to single-qubits to avoid additional complexity. Single-qubit operations can also be written in terms of three operators $X, Y$, and $Z$, where we consider only $X$ for the sake of example, which is sufficient for our purposes. Therefore, we assume that at time $t=0$ the rf signal is properly turned on to create gate $X$. In this case, immediately after this gate, the density matrix in the rotating system is:
\b \widetilde{\rho}_X=X\widetilde{\rho}_{th} X^{†}=\frac{1}{2} (I-\frac{\hbar\omega_0}{K_B T} I_y )  \label{eq:rox}\e
From which the density matrix in the laboratory system is:
\b \rho_X=U^{\dag} \widetilde{\rho}_{th} U=\frac{1}{2} (1-\frac{\hbar\omega_0}{K_B T}  [\sin(\omega_0 t) I_x+\cos(\omega_0 t) I_y ])\e
In this case the X-component of magnetization is:
\b M_x\propto\langle I_x \rangle=\textbf{tr}(I_x \rho)=-\frac{\hbar\omega_0}{4K_B T} sin(\omega_0 t)\e
Therefore the Fourier amplitude at the frequency $\omega_0$ is $\frac{\hbar\omega_0}{4K_BT}$. In the theory of  NMR quantum computers, the magnetic moment is measured in the laboratory system. But for now we temporarily assume that we measure the magnetic moment in the rotating system. In this case we have:
\b \widehat{M}_x \propto\langle I_x\rangle=\textbf{tr}(I_x\widehat{\rho}_X )=\textbf{tr}(I_x\frac{ 1}{2} (I-\frac{\hbar\omega_0}{K_B T} I_y )=0 \e
In this system, the magnetic moment in the Y direction is exactly equal to the magnetic moment in the X direction of the laboratory system:
\b \widehat{M}_y \propto\langle I_y\rangle=\textbf{tr}(I_y\widehat{\rho}_X )=\textbf{tr}(I_y\frac{ 1}{2} (I-\frac{\hbar\omega_0}{K_B T} I_y ))=\frac{\hbar\omega_0}{4K_B T} \e
That is why we assumed that the measurement was performed in a rotating device.
For this reason, we assumed that the measurement is performed in a rotating system. Therefore, to avoid quantum tomography operations, we work in a rotating system. We only need to keep in mind that the magnetization in the X direction in the laboratory system is the same as the magnetization in the Y direction in the rotating system and vice versa.

The above measurement theory was developed in the Copenhagen school. In fact, in the formula $[A]=\textbf{tr} (\rho A)$, which is the expected value of the observable $A$, quantum jumps are averaged:
\b [\widehat{A}]\equiv \Sigma_{i} w_i \langle \alpha^i |\widehat{A}| \alpha^i\rangle\e
where $w_i$ is the probability of jumping to the particular state $| \alpha^i\rangle$. Now let's see what modifications are made by the QSD theory for measurements in such quantum computers.

\section{Measurements in liquid NMR quantum computers with QSD theory}
\label{sec:QSD Measurements in NMR}
In the QSD interpretation, known as the decoherence interpretation, the measurement does not occur suddenly but over a period of time during which the state of the system changes according to the formula ~(\ref{eq:QSD0}). As we have said, the system first starts in a thermal state, then undergoes a quantum operation (here the operation X). Immediately after the quantum operation, which for convenience is taken as the origin of time, $t=0$, the density matrix in the rotating system becomes the equation ~(\ref{eq:rox}). We rewrite this density matrix as follows:
\b \widehat{\rho}_x(0)=\widehat{\rho}_{x1,1}|-\rangle\langle-|+\widehat{\rho}_{x1,2}|-\rangle\langle+|+\widehat{\rho}_{x2,1}|+\rangle\langle-|+\widehat{\rho}_{x2,2}|+\rangle\langle+|\e
where $\widehat{\rho}_{X m,n}$ is the $mn$ component of the ~(\ref{eq:rox}) matrix. The measurement process starts at $t=0$ and during that the states $|+\rangle $ and $|-\rangle$ evolve according to the formula ~(\ref{eq:QSD0}).
\b d|+,t\rangle =-\frac{i}{\hbar} H_o|+,t\rangle dt+\Sigma_j (\langle L_j^{\dag} \rangle L_j-\frac{1}{2} L_j^{\dag} L_j-\frac{1}{2}\langle L_j^{\dag} \rangle \langle L_j \rangle )|+,t\rangle dt+\Sigma_j(L_j-\langle  L_j \rangle)  |+,t\rangle dw \e
\b d|-,t\rangle =-\frac{i}{\hbar} H_o|-,t\rangle dt+\Sigma_j (\langle L_j^{\dag} \rangle L_j-\frac{1}{2} L_j^{\dag} L_j-\frac{1}{2}\langle L_j^{\dag} \rangle \langle L_j \rangle )|-,t\rangle dt+\Sigma_j(L_j-\langle  L_j \rangle)  |-,t\rangle dw \e
Now, by holding the state of the system at a desired instant, the density matrix at that instant is obtained. Then the magnetic moment, $M_y\propto \langle I_y\rangle=\textbf{tr}(I_y \rho)$, is obtained. Note that during the measurement the Hamiltonian of the system is $H_0$, because after the quantum operation, $B_1$ is turned off and during the measurement the field $B_0$ is still on.

Now we will look at the classification of measurements based on the intensity of the interaction between the measurement device and the system. To examine these different modes of measurement, we weight the contributions of the device and the environment in the formula ~(\ref{eq:QSD0}) with coefficients $\alpha$ and $\beta$:
\b |d\psi\rangle_{\alpha,\beta} =-\alpha\frac{i}{\hbar} H|\psi\rangle dt+\beta\Sigma_j (\langle L_j^{\dag} \rangle L_j-\frac{1}{2} L_j^{\dag} L_j-\frac{1}{2}\langle L_j^{\dag} \rangle \langle L_j \rangle )|\psi\rangle dt+\beta\Sigma_j(L_j-\langle  L_j \rangle)  |\psi\rangle dw \e
In this case, if $\alpha=1$ and $\beta=0$, the device is isolated, and if $\alpha=0$ and $\beta=1$, the device is wide open \cite{RN17}.

For numerical calculation, we need to know the typical values of the system parameters. The typical values of the parameters of NMR quantum computers are generally as follows: $B_0$ is about ten Tesla, from which the Larmor frequency $\omega_0$ is of the order of MHz (for example, for the hydrogen nucleus, it is five hundred MHz). T is the room temperature. The measurement time is about seconds, which we divide into about 500 parts in the numerical method.

Now we turn to the Lindblad operator. The measurement operation causes energy absorption and hence damping. Therefore, the Lindblad operator is the annihilation operator:
\b L=\left( \begin{array}{clcr} 0 & 1 \\ 0 & 0 \\ \end{array} \right)\e
The code of this program is included in the appendix \ref{app:maplecode} of the article. The necessary explanations about each part of this program, which is written in Maple, are given in the program itself.

\begin{figure}[tb]
\centering
\includegraphics[width=1\textwidth]{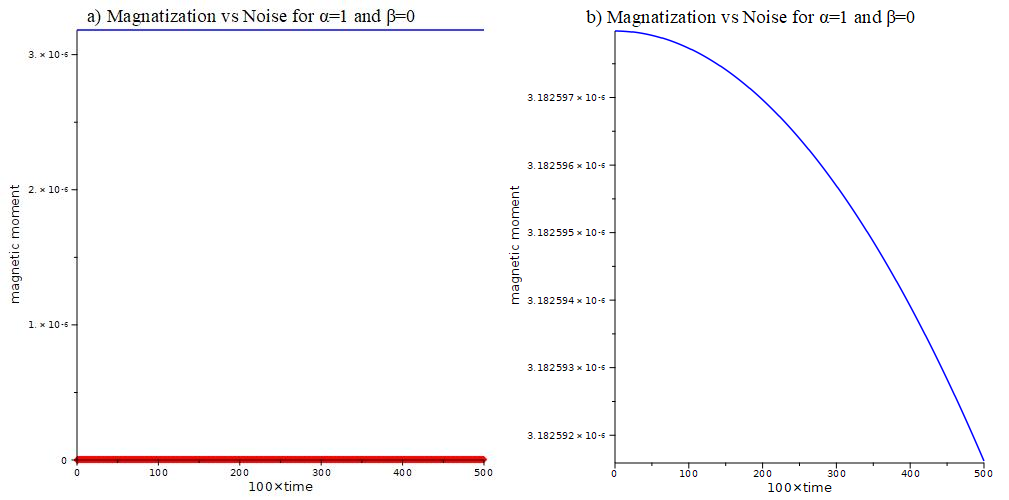}
\caption{\label{fig:beta=0}The measured magnetization of output of X gate for: a)signal and noise with $\alpha=1$ and $\beta=0$ b) only signal with $\alpha=1$ and $\beta=0$.}
\end{figure}

Let us examine the various interaction intensity. First, we consider the case where the system is isolated, that is, the measurement operation has no effect on the system. In this case, continuous measurement leads to the curve in Figure \ref{fig:beta=0}-a.

As can be seen from Figure \ref{fig:beta=0}-a, the magnetization is of the order of $10^{-6}$. In this case, as expected from the theory, the magnetization value is:

\b \widehat{M}_y \propto\langle I_y\rangle=\textbf{tr}(I_y\widehat{\rho}_X )=\frac{\hbar\omega_0}{4K_B T}=3.18\times10^{-6} \e

which is in good agreement with the value calculated from the QSD theory depicted in Figure\ref{fig:beta=0}-a. In order to have an estimate of the background noise, we also plot the noise. To estimate the noise, we plot the magnetization in the Y direction (Which in a rotating frame is equal to the magnetization in the X direction). But in this case,  since there is a large difference in the amplitudes of the noise and the signal, it is difficult to correctly identify the changes in the signal curve. Therefore, in Figure \ref{fig:beta=0}-b, we only plotted the signal so that we can better see these changes. As can be seen from the Figure \ref{fig:beta=0}-b, the magnetization decreases with time, which is due to the rounding error of the numerical calculations.

\begin{figure}[!]
\centering
\includegraphics[width=1\textwidth]{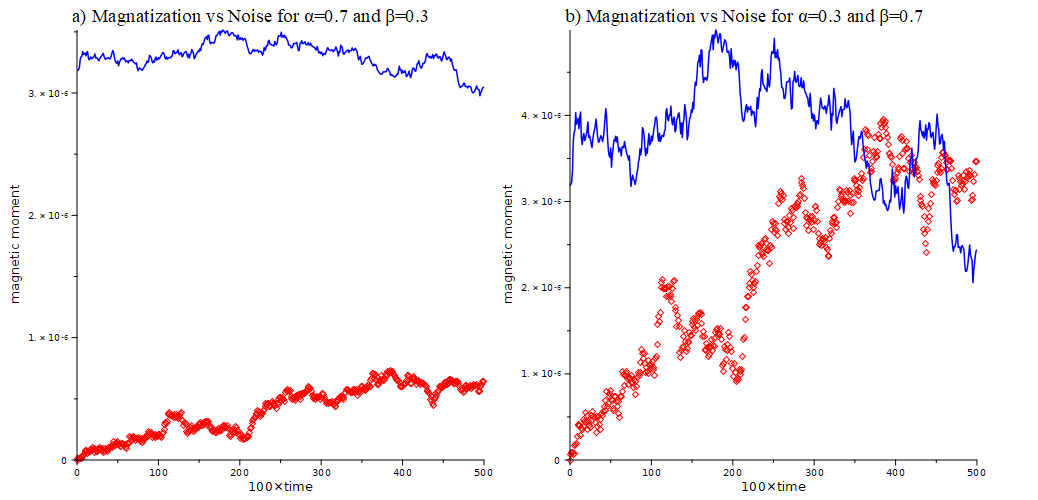}
\caption{\label{fig:beta=0.3_beta=0.7}The measured magnetization of output of X gate for: a)signal and noise with $\alpha=0.7$ and $\beta=0.3$ b) signal and noise with $\alpha=0.3$ and $\beta=0.7$.}
\end{figure}

Then we examine the case of $\alpha=0.7$ and $\beta=0.3$. With these coefficients, we expect the measurement effect to be not very large.  The result is Figure \ref{fig:beta=0.3_beta=0.7}-a.

This graph shows that, assuming weak interactions, extending the experiment time does not pose any particular problem. Now we consider the situation where the measurement is more robust. Let us assume $\alpha=0.3$ and $\beta=0.7$. In this case, the graph looks like Figure \ref{fig:beta=0.3_beta=0.7}-b.

\begin{figure}[!]
\centering
\includegraphics[width=1\textwidth]{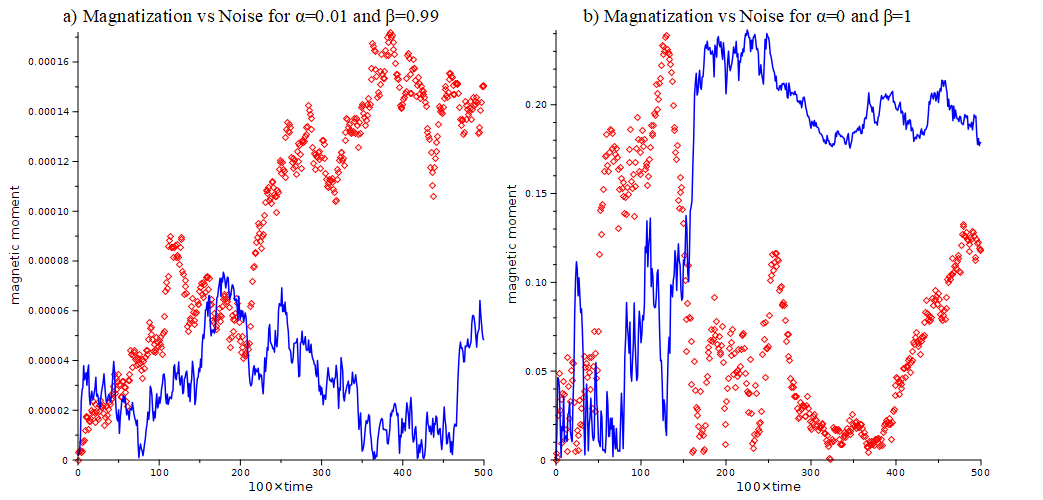}
\caption{\label{fig:beta=0.99_beta=1}The measured magnetization of output of X gate for: a)signal and noise with $\alpha=0.01$ and $\beta=0.99$ b) signal and noise with $\alpha=0$ and $\beta=1$.}
\end{figure}

We see that in this case the noise contribution increases. It can be clearly seen from the diagram that with increasing interaction intensity the noise contribution increases so that the accuracy of the results can be doubted. This means that in a situation where the noise is high, the measurement results cannot be counted on. To see this situation, we assume that the measurement is wide open, that is, $\alpha=0.01$ and $\beta=0.99$. The result will be as shown in Figure \ref{fig:beta=0.99_beta=1}-a.

As can be seen from the Figure \ref{fig:beta=0.99_beta=1}-a, for a while the signal is still stronger than the noise, but finally at some point the noise becomes comparable to the signal. Obviously, if the measurement has such a weight, the measurement duration must be very short (the simulated measurement duration is one second) for the measurement process to be meaningful.
In the fully wide open case, that is, $\alpha=0$ and $\beta=1$ , the result is as shown in Figure \ref{fig:beta=0.99_beta=1}-b.

\begin{figure}[!]
\centering
\includegraphics[width=0.5\textwidth]{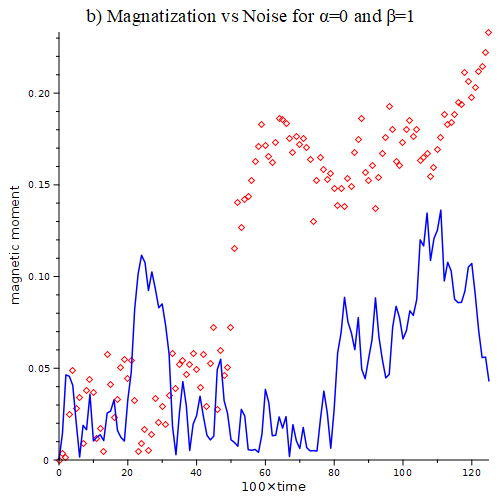}
\caption{\label{fig:beta=0.99_beta=1_zoom}The measured magnetization of output of X gate for signal and noise with $\alpha=0$ and $\beta=1$ for $t\in[0,0.25]$.}
\end{figure}

As can be seen from the Figure \ref{fig:beta=0.99_beta=1}-b, the greater the contribution of the measurement, i.e., the interference of the environment, the greater the fluctuation of the magnetization, so that it becomes comparable to the noise of the environment. To see the time interval that the noise has not yet dominated, we draw the first part of the last curve more broadly. The result is Figure \ref{fig:beta=0.99_beta=1_zoom}. We see that the noise and signal are mixed from the very beginning. Therefore, in this case the measurement is completely distorted and the result should be ignored. We postpone the conclusion and summary to the next section.

\section{Conclusions}
\label{sec:Conclu}
Quantum mechanics has always faced conceptual problems since its birth, some of which still remain. One of these problems is the measurement problem \cite{RN1,RN2}. The common interpretation is the Copenhagen interpretation. In the Copenhagen interpretation, it is not clear when the exact moment of measurement is. Also, in this interpretation, the device undergoes sudden jumps at the moment of measurement for which there is no dynamic equation. The jumps have always been mysterious. Competing interpretations have each tried to explain quantum jumps in some way. One of these competitors is the decoherence theory, which we applied to the measurement problem in NMR quantum computers in this article. In this interpretation, which is called QSD, the jumps are due to the statistical nature of the structural equations. The QSD theory claims that there is no jump. The jump is simply a solution to the Schrödinger stochastic differential equation.

In this article, we have not addressed the measurement-induced jump. Rather, by applying the QSD theory to measurements in NMR quantum computers, we have observed that according to this theory, if the measurement is not to cause severe disturbance in the system, either the measurement time must be short or the interaction must not be strong. This feature is important because the signal must be distinguishable from the noise. Shortening the measurement time has its limitations. Because by shortening the measurement time interval, the received energy decreases and the signal may not be distinguishable from the noise. Therefore, we pay attention to the option of reducing the interaction strength. The interaction strength can be controlled with magnetization measurement coils. On the one hand, these coils are used for two purposes. First, to create the rf frequency and second, to measure the magnetization. So here we are dealing with a measurement equipment engineering problem. In fact, to create a specific $\omega_{1}$, the amplitude of the rf field must be freely changed, so the dimensions of the coil generating the rf frequency may be large, and if the same coils are used to measure magnetization, it may cause a large $\beta$ and, as a result, the measurement will be distorted. Therefore, our suggestion is to use two separate coils, one for creating the rf frequency and the other for measuring magnetization, in order to better control the measuring device. We have postponed the topic of mutations in QSD theory to our next article.

\vspace{0.5cm} \noindent {\bf{Acknowledgements}}: The authors sincerely thank Dr. Ramzan Rezeyan, Department of Mathematics, Islamic Azad University, Nour Branch, for very useful and valuable discussions.
 \vspace{0.5cm}

\begin{appendix}

\section{The Maple code of the program}
\label{app:maplecode}
The Maple code to calculate the magnetization after applying the gate X as well as the magnetization of the background noise is given below.
\\ \\

\begin{Maple Normal}
{$ \displaystyle \mathit{restart}  $}
\end{Maple Normal}
\begin{Maple Normal}
{$ \displaystyle \mathit{with} (\mathit{ScientificConstants})\colon  $}
\end{Maple Normal}
\begin{Maple Normal}
{$ \mathit{with} (\mathit{LinearAlgebra})\colon  $}
\end{Maple Normal}
\begin{Maple Normal}
{$ \displaystyle \mathit{with} (\mathit{Physics})\colon  $}
\end{Maple Normal}
\begin{Maple Normal}
{$ \displaystyle \mathit{with} (\mathit{SignalProcessing})\colon  $}
\end{Maple Normal}
\begin{Maple Normal}
{$ \displaystyle \mathit{with} (\mathit{Statistics})\colon  $}
\end{Maple Normal}

\begin{Maple Normal}
{$ \displaystyle \esnum \texttt{ The alfa and beta coefficient that control the strength of measurement process.} $}
\end{Maple Normal}

\begin{Maple Normal}
{$ \alpha \coloneqq  0.7\colon  $}
{$ \beta \coloneqq  0.3\colon  $}
\end{Maple Normal}
\begin{Maple Normal}
{$ \displaystyle \esnum \mathit{\,Planck's\,constant\,and\,gases's\,constant\,}  $}
\end{Maple Normal}
\begin{Maple Normal}
{$ \displaystyle \mathrm{hbar}\coloneqq \mathit{GetValue} (\mathit{Constant} (\esapos \mathrm{hbar}\esapos))\colon K [B]\coloneqq \mathit{GetValue} (\mathit{Constant} (\esapos k \esapos))\colon  $}
\end{Maple Normal}
\begin{Maple Normal}
{$ \displaystyle T \coloneqq 300\colon \esnum \,\mathit{The} \mathit{tempreture} \mathit{of} \mathit{the} \mathit{system}  $}
\end{Maple Normal}
\begin{Maple Normal}
{$ \mathrm{omega}[0]\coloneqq 500\cdot 10^{6}\colon  $}
{$ \esnum \texttt{ the Larmor frequency(=}\gamma \cdot \texttt{B0)}  $}
\end{Maple Normal}
\begin{Maple Normal}
{$ \mathit{piriod} \coloneqq \mathit{evalf} (\frac{2\cdot \mathrm{Pi}}{500\cdot 10^{6}})\colon \esnum \mathit{\,The\,piriod\,of\,Larmor\,frequency\,}  $}
\end{Maple Normal}
\begin{Maple Normal}
{$ \frac{\mathrm{hbar}\cdot \mathrm{omega}[0]}{4\cdot K [B]\cdot T}\colon  $}
{$ \esnum \mathit{\,The\,ratio\,of\,spin\,jump\,energy\,to\,thermal\,energy\,of\,nucleus\,}  $}
\end{Maple Normal}
\begin{Maple Normal}
{$ \displaystyle \mathit{SigmaX} \coloneqq \left[\begin{array}{cc}
0 & 1 
\\
 1 & 0 
\end{array}\right]\colon \mathit{SigmaY} \coloneqq \left[\begin{array}{cc}
0 & -I  
\\
 I  & 0 
\end{array}\right]\colon \mathit{SigmaZ} \coloneqq \left[\begin{array}{cc}
1 & 0 
\\
 0 & -1 
\end{array}\right]\colon \esnum \mathit{\,the\,Pauli\,matrises\,}  $}
\end{Maple Normal}

\begin{Maple Normal}
{$ \displaystyle \mathit{Ix} \coloneqq  0.5\cdot \mathit{SigmaX} \colon \mathit{Iy} \coloneqq  0.5\cdot \mathit{SigmaY} \colon \mathit{Iz} \coloneqq  0.5\cdot \mathit{SigmaZ} \colon \esnum \mathit{The\,I\,matrises\,}  $}
\end{Maple Normal}
\begin{Maple Normal}
{$ \displaystyle \esnum \phi :=\mathit{evalf} (\frac{\mathrm{Pi}}{2}) $}
\end{Maple Normal}
\begin{Maple Normal}
{$ \displaystyle \mathit{100000} \coloneqq 100\cdot 10^{3}\colon  $}
\end{Maple Normal}
\begin{Maple Normal}
{$ X \coloneqq \frac{1}{\sqrt{2}}.\left[\begin{array}{cc}
1 & -I  
\\
 -I  & 1 
\end{array}\right]\colon  $}
{$ Y \coloneqq \frac{1}{\sqrt{2}}.\left[\begin{array}{cc}
1 & -1 
\\
 1 & 1 
\end{array}\right]\colon  $}
{$ Z \coloneqq \frac{1}{\sqrt{2}}.\left[\begin{array}{cc}
1-I  & 0 
\\
 0 & 1+I  
\end{array}\right]\colon \esnum \mathit{The\,X,\,Y\,and\,Z\,gates.\,}  $}
\end{Maple Normal}
\begin{Maple Normal}
{$ \mathit{H0} \coloneqq -\mathrm{hbar}\cdot \mathrm{omega}[0]\cdot \mathit{Iz} \colon \esnum \mathit{\,The\,Hamiltonian\,without\,the\,rf\,field\,}  $}
\end{Maple Normal}
\begin{Maple Normal}
{$ \displaystyle \esnum \texttt{ The Hamiltonian in rotating wave approximation.}  $}
\end{Maple Normal}
\begin{Maple Normal}
{$ \displaystyle \esnum \texttt{ In this aproximation we have:}\omega \texttt{=}\omega \texttt{rf=}\omega \texttt{0, where}~\omega \texttt{ is the rotating frame angular} $}

{$ \displaystyle \esnum  \texttt{velocity, where}~\omega \texttt{ is the rotating frame angular velocity,} $}
\end{Maple Normal}

\begin{Maple Normal}
{$ \displaystyle \esnum ~\omega\texttt{rf the radio field angular ferequency and}  ~\omega \texttt{0 the Larmor frequency.}  $}
\end{Maple Normal}

\begin{Maple Normal}
{$ \displaystyle \esnum~\phi\texttt{ is the rf field initial phase.} $}
\end{Maple Normal}

\begin{Maple Normal}
{$ \displaystyle \esnum ~H:= \mathrm{hbar}\cdot \mathit{100000} .(\cos (\phi)\cdot \mathit{Ix} +\sin (\phi)\cdot \mathit{Iy})\colon  $}
\end{Maple Normal}

\begin{Maple Normal}
{$ \displaystyle \esnum \texttt{The density matrix of a single nucleus without the rf field at thermal} $}

{$ \displaystyle \esnum \texttt{equilibrium in therotating frame.} $}
\end{Maple Normal}

\begin{Maple Normal}
{$ \displaystyle \rho [\mathit{th}]\coloneqq  0.5\cdot (\mathit{IdentityMatrix} (2)-\frac{\mathit{H0}}{K [B]\cdot T})\colon  $}
\end{Maple Normal}

\begin{Maple Normal}
{$ \displaystyle \esnum \texttt{the density matrix after X operator inrotating frame.} $}
\end{Maple Normal}
 \begin{Maple Normal}
{$ \displaystyle \rho [\mathit{afterX}]\coloneqq X \cdot \rho [\mathit{th}].\mathit{Transpose} (\mathit{conjugate} (X))\colon  $}
\end{Maple Normal}
\begin{Maple Normal}
{$ N \coloneqq 500\colon  $}
{$ \esnum \mathit{\,the\,number\,of\,division\,of\,measurment\,time\,}  $}
\end{Maple Normal}
\begin{Maple Normal}
{$ \displaystyle t \coloneqq 1\colon \esnum \mathit{The\,measurment\,time\,duration\,}  $}
\end{Maple Normal}
\begin{Maple Normal}

{$ \mathit{dt} \coloneqq \frac{t}{N}\colon \esnum \mathit{\,The\,time\,inceament\,}  $}
\end{Maple Normal}
\begin{Maple Normal}
{$ \displaystyle L \coloneqq \left[\begin{array}{cc}
0 & 1 
\\
 0 & 0 
\end{array}\right]\colon \esnum \mathit{\,Lindblad\,operator\,for\,of\,the\,system\,}  $}
\end{Maple Normal}
\begin{Maple Normal}
{$ \displaystyle \mathit{Lt} \coloneqq \mathit{Transpose} (\mathit{conjugate} (L))\colon \esnum \mathit{\,The\,transpose\,of\,L\,}  $}
\end{Maple Normal}
\begin{Maple Normal}
{$ \displaystyle \mathit{LtL} \coloneqq \mathit{Lt} \cdot L \colon  $}
\end{Maple Normal}
\begin{Maple Normal}
{$ \mathit{Sii1} [0]\coloneqq \mathit{Matrix} (2,1)\colon  $}
{$ \esnum \mathit{\,the\,base\,state\,No\,1\,}  $}
\end{Maple Normal}
\begin{Maple Normal}
{$ \mathit{Sii1} [0][2,1]\coloneqq 1\colon \esnum \mathit{\,\,the\,up\,state\,}  $}
\end{Maple Normal}
\begin{Maple Normal}
{$ \displaystyle \mathit{Sii1} [0]\colon  $}
\end{Maple Normal}
\begin{Maple Normal}
{$ \displaystyle \mathit{Siit1} [0]\coloneqq \mathit{Matrix} (1,2)\colon \esnum \mathit{\,transpose\,of\,the\,base\,state\,}  $}
\end{Maple Normal}
\begin{Maple Normal}
{$ \mathit{dSii1} [0]\coloneqq \mathit{Matrix} (2,1)\colon \esnum \mathit{\,differential\,of\,state\,}  $}
\end{Maple Normal}
\begin{Maple Normal}
{$ \mathit{Sii2} [0]\coloneqq \mathit{Matrix} (2,1)\colon  $}
{$ \esnum \mathit{\,the\,base\,state\,No\,2\,}  $}
\end{Maple Normal}
\begin{Maple Normal}
{$ \displaystyle \mathit{Sii2} [0][1,1]\coloneqq 1\colon \esnum \mathit{\,the\,down\,state\,}  $}
\end{Maple Normal}
\begin{Maple Normal}
{$ \displaystyle \mathit{Sii2} [0]\colon  $}
\end{Maple Normal}
\begin{Maple Normal}
{$ \displaystyle \mathit{Siit2} [0]\coloneqq \mathit{Matrix} (1,2)\colon \esnum \mathit{\,transpose\,of\,the\,state\,}  $}
\end{Maple Normal}
\begin{Maple Normal}
{$ \displaystyle \mathit{dSii2} [0]\coloneqq \mathit{Matrix} (2,1)\colon \esnum \mathit{\,differential\,of\,state\,}  $}
\end{Maple Normal}
\begin{Maple Normal}

{$ \displaystyle \esnum \texttt{The real and imaginary parts of Gaussian noises with mean equel to} $}

{$ \displaystyle \esnum \texttt{ zero and standard diviation} \sqrt{\mathit{dt}} $}
\end{Maple Normal}
\begin{Maple Normal}
{$ \displaystyle \mathit{c11} \coloneqq \mathit{Array} (1..(N +1),{\esapos \mathit{datatype} \esapos}={\esapos \mathit{float} \esapos}[8],{\esapos \mathit{order} \esapos}={\esapos \textit{C\_order} \esapos})\colon  $}
\end{Maple Normal}
\begin{Maple Normal}
{$ \displaystyle \mathit{GenerateGaussian} (N +1,0,\sqrt{\mathit{dt}},\mathit{Complex} ,\esapos \mathit{container} \esapos =\mathit{c11})\colon  $}
\end{Maple Normal}
\begin{Maple Normal}
{$ \displaystyle \mathit{c12} \coloneqq \mathit{Array} (1..(N +1),{\esapos \mathit{datatype} \esapos}={\esapos \mathit{float} \esapos}[8],{\esapos \mathit{order} \esapos}={\esapos \textit{C\_order} \esapos})\colon  $}
\end{Maple Normal}
\begin{Maple Normal}
{$ \displaystyle \mathit{GenerateGaussian} (N +1,0,\sqrt{\mathit{dt}},\mathit{Complex} ,\esapos \mathit{container} \esapos =\mathit{c12})\colon  $}
\end{Maple Normal}
\begin{Maple Normal}
{$ \displaystyle \mathit{Mx} \coloneqq \mathit{Seq} (0..N ,{\esapos \mathit{order} \esapos}={\esapos \textit{C\_order} \esapos})\colon \esnum \texttt{ magnatization along x at time i}\times \mathit{dt}  $}
\end{Maple Normal}
\begin{Maple Normal}
{$ \displaystyle \mathit{My} \coloneqq \mathit{Seq} (0..N ,{\esapos \mathit{order} \esapos}={\esapos \textit{C\_order} \esapos})\colon \esnum \texttt{ magnatization along y at time i}\times \mathit{dt}  $}
\end{Maple Normal}
\begin{Maple Normal}
{$ \displaystyle \rho [\mathit{dm}][0]\coloneqq \mathit{Matrix} (2,2)\colon \esnum \texttt{ the density matrix during measurment process at time i}\times \mathit{dt}  $}
\end{Maple Normal}

\begin{Maple Normal}
{$ \displaystyle \esnum \texttt{ the density matrix during measurment process at time i}\times \mathit{dt}  $}
\end{Maple Normal}
\begin{Maple Normal}
{$ \displaystyle \rho [\mathit{dm}][0]\coloneqq \mathit{Matrix} (2,2)\colon   $}
\end{Maple Normal}
\begin{Maple Normal}
{$ \displaystyle \esnum\texttt{ We then compute the state of the system at any time t and then compute  } $}
\end{Maple Normal}
\begin{Maple Normal}
{$ \displaystyle \esnum\texttt{ the density matrix and the magnatization of the system at this time. } $}
\end{Maple Normal}

\begin{Maple Normal}
  $\boldsymbol{for}~i \boldsymbol{~from}~0 \boldsymbol{~by} \boldsymbol{~to} ~N  \boldsymbol{~do}$
\\
  $Siit1[i]\coloneqq Transpose (conjugate (Sii1[i]));LtMean1 \coloneqq Siit1[i]\cdot Lt\cdot Sii1[i];LMean1\coloneqq Siit1[i]\cdot L \cdot Sii1[i];$
\\
 {$\mathit{Siit2} [i]\coloneqq \mathit{Transpose} (\mathit{conjugate} (\mathit{Sii2} [i]));\mathit{LtMean2} \coloneqq \mathit{Siit2} [i]\cdot \mathit{Lt} \cdot \mathit{Sii2} $} {$[i];\mathit{LMean2} \coloneqq \mathit{Siit2} [i]\cdot L \cdot \mathit{Sii2} [i];$}

{$ \rho [\mathit{dm}][i]\coloneqq \rho [\mathit{afterX}][1,1]\cdot \mathit{Sii1} [i]\cdot \mathit{Siit1} [i]+\rho [\mathit{afterX}][1,2]\cdot \mathit{Sii1} [i]\cdot $} {$\mathit{Siit2} [i]+\rho [\mathit{afterX}][2,1]\cdot \mathit{Sii2} [i]\cdot \mathit{Siit1} [i]+\rho [\mathit{afterX}][2,2]\cdot \mathit{Sii2} [i]\cdot \mathit{Siit2}$} {$[i];\mathit{Mx} [i]\coloneqq \mathit{Trace} (\rho [\mathit{dm}][i]\cdot \mathit{Ix});\mathit{My} [i]\coloneqq \mathit{Trace} (\rho [\mathit{dm}][i]\cdot \mathit{Iy});$}

 {$\mathit{dSii1} [i +1]\coloneqq -\alpha \cdot I .(\frac{\mathit{H0}}{\mathrm{hbar}}\cdot \mathit{Sii1} [i])\cdot \mathit{dt} +\beta .(\mathit{LtMean1} [1,1]\cdot L - 0.5\cdot $} {$(\mathit{LtL})- 0.5\cdot (\mathit{LtMean1} [1,1])\cdot (\mathit{LMean1} [1,1])\cdot \mathit{IdentityMatrix} (2))\cdot \mathit{Sii1} [i]\cdot \mathit{dt} +\beta .(L $} {$-\mathit{LMean1} [1,1]\cdot \mathit{IdentityMatrix} (2))\cdot \mathit{Sii1} [i]\cdot (\mathit{c11} [i +1]+I \cdot \mathit{c12} [i +1]);$}
\\
 {$;\mathit{Sii1} [i +1]\coloneqq \frac{(\mathit{Sii1} [i]+\mathit{dSii1} [i +1])}{\mathit{Norm} (\mathit{Sii1} [i]+\mathit{dSii1} [i +1],2)};$}

 {$\mathit{dSii1} [i +1]\coloneqq -\alpha \cdot I .(\frac{\mathit{H0}}{\mathrm{hbar}}\cdot \mathit{Sii1} [i])\cdot \mathit{dt} +\beta .(\mathit{LtMean1} [1,1]\cdot L - 0.5\cdot $} {$(\mathit{LtL})- 0.5\cdot (\mathit{LtMean1} [1,1])\cdot (\mathit{LMean1} [1,1])\cdot \mathit{IdentityMatrix} (2))\cdot \mathit{Sii1} [i]\cdot \mathit{dt} +\beta .(L $} {$-\mathit{LMean1} [1,1]\cdot \mathit{IdentityMatrix} (2))\cdot \mathit{Sii1} [i]\cdot (\mathit{c11} [i +1]+I \cdot \mathit{c12} [i +1]);$}
\\
 {$;\mathit{Sii1} [i +1]\coloneqq \frac{(\mathit{Sii1} [i]+\mathit{dSii1} [i +1])}{\mathit{Norm} (\mathit{Sii1} [i]+\mathit{dSii1} [i +1],2)};$}

{$\mathit{dSii2} [i +1]\coloneqq -\alpha \cdot I .(\frac{\mathit{H0}}{\mathrm{hbar}}\cdot \mathit{Sii2} [i])\cdot \mathit{dt} +\beta .(\mathit{LtMean1} [1,1]\cdot L - 0.5\cdot $} {$(\mathit{LtL})- 0.5\cdot (\mathit{LtMean1} [1,1])\cdot (\mathit{LMean1} [1,1])\cdot \mathit{IdentityMatrix} (2))\cdot \mathit{Sii2} [i]\cdot \mathit{dt} +\beta .(L $} {$-\mathit{LMean1} [1,1]\cdot \mathit{IdentityMatrix} (2))\cdot \mathit{Sii2} [i]\cdot (\mathit{c11} [i +1]+I \cdot \mathit{c12} [i +1]);$}
\\
 {$;\mathit{Sii2} [i +1]\coloneqq \frac{(\mathit{Sii2} [i]+\mathit{dSii2} [i +1])}{\mathit{Norm} (\mathit{Sii2} [i]+\mathit{dSii2} [i +1],2)};$}

 {$\boldsymbol{\mathrm{end}}\boldsymbol{\mathrm{do}}\colon  $}
\end{Maple Normal}
 \begin{Maple Normal}
{$ \displaystyle \mathit{with} (\mathit{plots})\colon  $}
\end{Maple Normal}
\begin{Maple Normal}
{$ \displaystyle \mathit{plot} ({\{[\mathit{seq} ([k ,\mathrm{abs}(\Re (\mathit{Mx} [k]))],k =0..N)],[\mathit{seq} ([k ,\mathrm{abs}(\Re (\mathit{My} [k]))],k =0..N)],\}},\mathit{style} =[\mathit{point} ,\mathit{line}],$}
 {$ \displaystyle \mathit{color} =[\text{``Red"},\text{``Blue"}], \mathit{titlefont} =[\text{``ARIAL"},15],\mathit{labels} =[\text{``100$\times $time"},\text{``magnetic moment"}],$}
 
{$ \displaystyle \mathit{labeldirections} =[\text{``horizontal"},\text{``vertical"}],\mathit{labelfont} =[\text{``HELVETICA"},10],\mathit{linestyle} =[\mathit{solid} ,\mathit{solid}],$}
{$ \displaystyle \mathit{axesfont} =[\text{``HELVETICA"},\text{``ROMAN"},8],\mathit{legendstyle} =[\mathit{font} =[\text{``HELVETICA"},9],\mathit{location} =\mathit{right}])\colon  $}
\end{Maple Normal}

\end{appendix}

\bibliography{NMRQc}

%\begin{thebibliography}{NMR QC measurement problem}

%\end{thebibliography}

\end{document}